\documentclass[aps,prl,twocolumn,floatfix,showpacs]{revtex4}
\usepackage{graphics}

\begin{document}

\title{Self-Doping of Gold Chains on Silicon:  A New Structural Model for Si(111)5$\times$2-Au}
\author{Steven C. Erwin}
\affiliation{Center for Computational Materials Science, Naval Research Laboratory,
Washington, D.C. 20375}
\date{\today}

\begin{abstract}
A new structural model for the Si(111)5$\times$2-Au reconstruction is
proposed and analyzed using first-principles calculations.  The basic
model consists of a ``double honeycomb chain'' decorated by Si
adatoms. The 5$\times$1 periodicity of the honeycomb chains is doubled
by the presence of a half-occupied row of Si atoms that partially
rebonds the chains.  Additional adatoms supply electrons that dope the
parent band structure and stabilize the period doubling; the optimal
doping corresponds to one adatom per four 5$\times$2 cells, in
agreement with experiment. All the main features observed in scanning
tunneling microscopy and photoemission are well
reproduced.
\end{abstract}

\pacs{68.43.Bc,73.20.At,68.35.Bs,81.07.Vb}
\maketitle

Physical realizations of a one-dimensional metal are rare, in part
because they may be preempted by a metal-insulator Peierls transition
\cite{himpsel_journal_of_electron_spectroscopy_and_related_phenomena_2002a}. An
escape clause is available for metallic chains adsorbed on rigid
substrates, however, since the energy penalty for the pairing
distortion may be prohibitively high.  For example, when gold is
adsorbed on silicon a variety of chain-like structures are
formed, some with unusual electronic properties suggestive of a
one-dimensional metal
\cite{segovia_nature_1999a,losio_phys_rev_lett_2001a}. Photoemission
data from the vicinal surfaces Si(553)-Au and Si(557)-Au reveal
fractionally filled bands with strongly one-dimensional
character, which are believed to originate from Au ``chains'' just one
atom wide \cite{crain_phys_rev_lett_2003a}.  Structural models for
Si(557)-Au were recently proposed based on total-energy calculations
\cite{sanchez-portal_phys_rev_b_2002a} and x-ray data
\cite{robinson_phys_rev_lett_2002a}.

Even more widely studied is the parent flat surface, Si(111)-Au. First
reported over twenty-five years ago \cite{lipson_j_phys_c_1974a,
lelay_surf_sci_1977a}, the Si(111)5$\times$2-Au reconstruction
has been extensively characterized by low-energy electron diffraction
\cite{bishop_british_journal_of_applied_physics_1969a,lelay_surf_sci_1977a},
scanning tunneling microscopy (STM)
\cite{baski_phys_rev_b_1990a,omahony_surf_sci_1992a,omahony_phys_rev_b_1994a}, x-ray
diffraction \cite{schamper_phys_rev_b_1991a}, 
reflectance anisotropy spectroscopy \cite{power_phys_rev_b_1997a},
angle-resolved
photoemission
\cite{collins_surf_sci_1995a,altmann_phys_rev_b_2001a}, 
inverse photoemission \cite{hill_appl_surf_sci_1998a},
and core-level spectroscopy
\cite{zhang_phys_rev_b_2002b}. 
Despite this scrutiny, its structure remains
unknown.  The data provide several constraints on any model of
Si(111)5$\times$2-Au. (1) STM shows the surface to be decorated by
bright protrusions with apparent height $\sim$1.5 \AA~and whose
coverage, although variable, has a preferred value of one per
5$\times$8 supercell \cite{bennewitz_nanotechnology_2002a}. (2) Away
from the protrusions, STM images show a ``Y''-shaped feature whose
orientation is determined by the underlying lattice
\cite{omahony_surf_sci_1992a,omahony_phys_rev_b_1994a,power_phys_rev_b_1997a}.
(3) Photoemission finds a strong
surface band beginning at the the 5$\times$2 zone boundary and
dispersing downward toward the 5$\times$1 zone boundary
\cite{altmann_phys_rev_b_2001a}. (4) The nature of this band changes
from one-dimensional at the top of the band to two-dimensional at its
bottom \cite{losio_phys_rev_lett_2000a}.

In this Letter a new model is proposed for Si(111)5$\times$2-Au that
explains all of these observed features. The model is related to those
proposed earlier for Si(557)-Au, suggesting that all of the
reconstructions formed by adsorption of Au on Si form a family.  It is
also closely related to the ``honeycomb chain-channel'' (HCC) model
now widely accepted as the structure of the adsorbate-induced induced
Si(111)3$\times$1 and Si(111)3$\times$2 reconstructions
\cite{erwin_phys_rev_lett_1998a}, and thus helps to unify a wide class of
adsorbate-induced reconstructions based on Si(111) and its 
vicinals. The present model achieves its stability through
an unusual ``self-doping'' mechanism that may be relevant to other
Au-induced Si reconstructions.

\begin{figure}[b]
\resizebox{6.75cm}{!}{\includegraphics{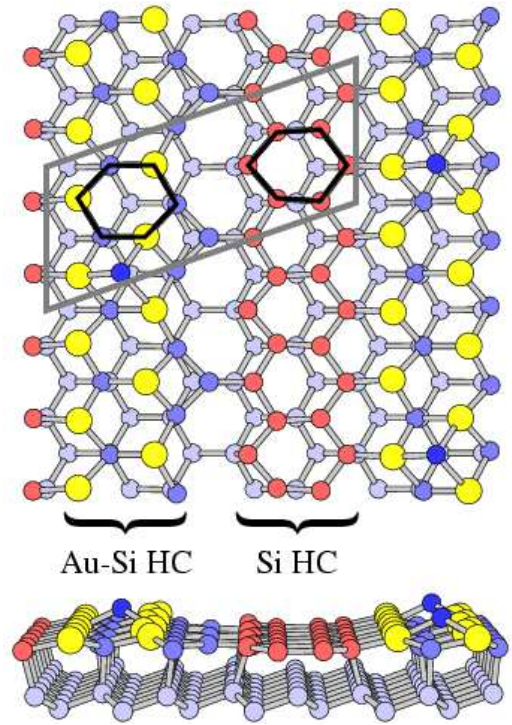}}
\caption{(color) Proposed ``double honeycomb chain'' structure of Si(111)5$\times$2-Au. Large circles are
Au, small circles are Si. The elementary 5$\times$2 unit cell is
outlined.  Each unit cell contains two honeycomb chains (HC)
based on the outlined hexagons, one of alternating Au and Si
atoms, the other of all Si.  Three additional Si adatoms, with
5$\times$4 periodicity, are also shown (see discussion). }
\end{figure}

The model is shown in Fig.~1. The reconstruction occurs purely in the
surface layer, and has the basic structure of a ``double honeycomb
chain'' (DHC) with underlying 5$\times$1 periodicity. One chain is
formed by hexagons of alternating Au and Si atoms, and one by hexagons
of all Si (as in the HCC model). The outer Si atoms of the Au-Si chain
are too far from the Si atoms in the Si chain to bond directly.
As a result, two variants of the basic model are plausible. (i) The
insertion of an additional ``rebonding'' row of Si atoms can bridge the gap
between the chains, but only at
the cost of overcoordinating the Si atoms in the Au-Si chain. This
will be called the ``5$\times$1 variant.''  (ii)
Removing every other of these rebonding Si atoms (as shown in Fig.~1)
relieves the overcoordination but
now leaves dangling bonds in the Si chain; this will be called the
``5$\times$2 variant.''  These two variants are
very close energetically.  It will be shown below that the
presence of additional Si, occurring as adatoms, acts to dope the
5$\times$2 variant with electrons and thereby to reduce its surface
energy relative to the 5$\times$1 variant.  The optimal doping level
occurs for one adatom per four 5$\times$2 cells, identical to 
the observed equilibrium adatom coverage.

First-principles total-energy calculations were used to determine the
equilibrium geometries and relative surface energies of the basic
model and its variants.  The calculations were performed in a slab
geometry with up to six layers of Si plus the reconstructed surface
layer and a vacuum region of 8 \AA. All atomic positions were relaxed
until the total energy changed by less than 1 meV per 5$\times$2 cell;
the bottom Si layer was passivated by hydrogen and held fixed. Total energies and
forces were calculated within the generalized-gradient approximation
to density-functional theory using projector-augmented-wave
potentials, as implemented in {\sc vasp}
\cite{kresse_phys_rev_b_1993a,kresse_phys_rev_b_1996a}.  The 
plane-wave cutoff (180 eV) and sampling (2$\times$4) of the surface
Brillouin zone were sufficient to converge relative surface energies
to within 1 meV/\AA$^2$, adequate for the comparisons
presented below. All the models considered here have equal Au
coverage, and hence the relative surface energies (calculated as in
Ref.~\onlinecite{erwin_phys_rev_lett_1998a}) do not require
a choice of Au chemical potential.  STM images were simulated using
the method of Tersoff and Hamann \cite{tersoff_phys_rev_b_1985a}.

Since its discovery, many structural models have been
proposed for Si(111)5$\times$2-Au.  Most of the early ones are not
compatible with newer STM data, but two later models---from Marks and
Plass (MP) \cite{marks_phys_rev_lett_1995a} and Hasegawa, Hosaka, and Hosoki (HHH)
\cite{hasegawa_surf_sci_1996a}---have recently been studied theoretically
\cite{kang_surf_sci_2003a}.  Neither was found to be consistent 
with STM or ARPES data, despite being locally stable with nearly equal
surface energies (to within 0.1 meV/\AA$^2$).  The 5$\times$2 DHC
model proposed here is more stable than the MP and HHH models by 20
meV/\AA$^2$, or 2.6 eV per 5$\times$2 cell.  This energy difference is
sufficiently large to rule out the MP and HHH models on energetic
grounds alone.

\begin{figure}[b]
\resizebox{6.0cm}{!}{\includegraphics{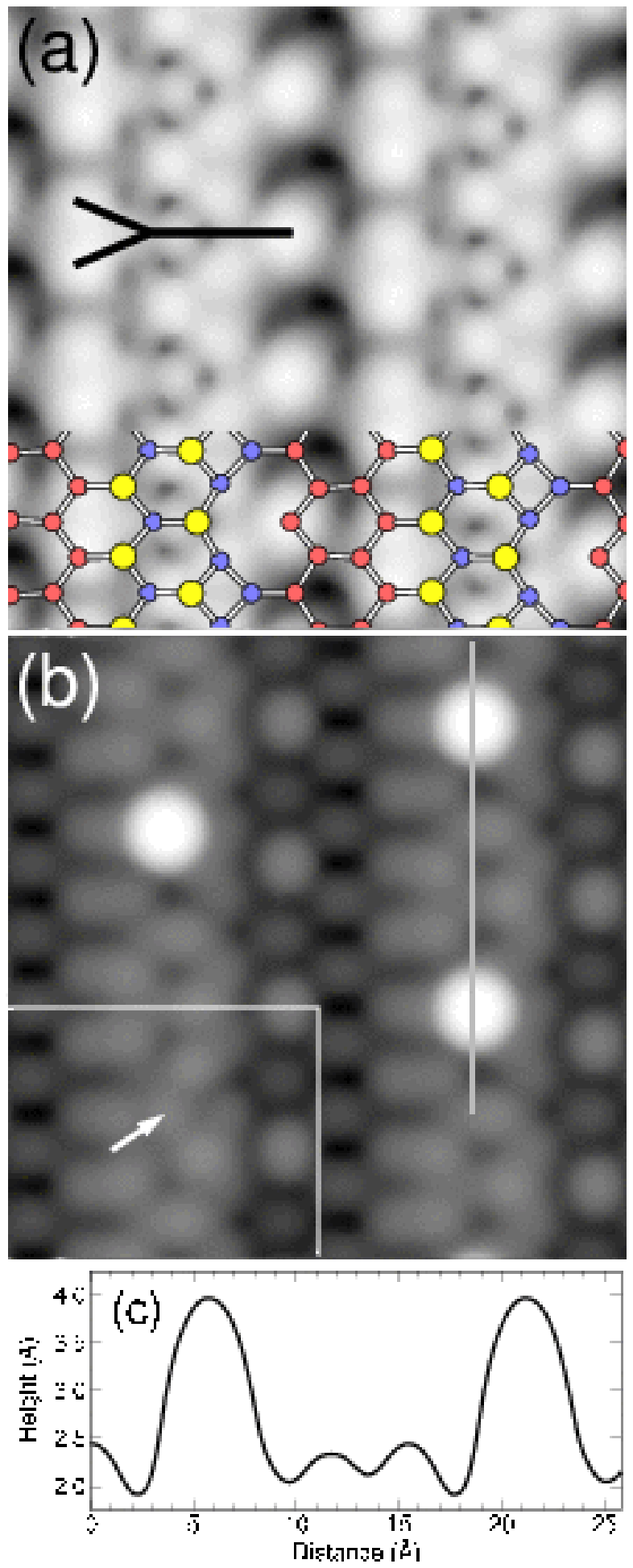}}
\caption{(color) Simulated filled-state STM images for double honeycomb chain model. (a) 
Image for 5$\times$2 model (sample bias $-$0.8 V), showing
``Y''-shaped feature observed in
Ref.~\onlinecite{omahony_surf_sci_1992a}. (b) Same surface, with
adsorbed Si atoms as in Fig.~1 (sample bias $-$2.0 V), showing bright
protrusions observed in
Ref.~\onlinecite{bennewitz_nanotechnology_2002a}. Inset: simulated
image from an adsorbed Au atom at the same in-plane location (marked
by arrow). (c) Linescan through two bright protrusions, as marked in
panel (b).}
\end{figure}

Within the family of the two DHC variants described above (5$\times$1 versus
5$\times$2, each with different Si adatom coverages) the relative
surface energy changes are very much smaller, of order 1 meV/\AA$^2$. This
is near the limiting precision for DFT surface energies, suggesting
the need for a simpler model to describe, for example, the variation
of surface energy with Si adatom coverage. In the following, this
model is developed by analyzing within DFT the role of the
adatoms. This requires demonstrating, first, that the observed bright
protrusions in STM are, in fact, Si adatoms and not something else;
and second, that the adatoms indeed supply electrons, which then dope
the parent band structure.

Away from the protrusions, filled-state STM images show a series of
side-by-side ``Y''-shaped features with 5$\times$2 periodicity
\cite{omahony_surf_sci_1992a,omahony_phys_rev_b_1994a,power_phys_rev_b_1997a}.
A simulated STM image for the 5$\times$2 DHC model with no adatoms is
shown in Fig.~2(a).  A similar feature ``Y''-shaped feature is found
with paired ``arms,'' a single ``tail,'' and the same crystallographic
orientation as found experimentally.

To identify the bright protrusions seen in STM images, both Si and Au
adatoms were considered as potential candidates.  Binding energies for
individual adatoms were calculated for eight possible sites on the
undecorated 5$\times$2 surface. For Si the most favorable site is at
the center of a Au-Si hexagon, as shown in Fig. 1, where the binding
energy is 4.6 eV.  This is much larger than the next best site (by 0.7
eV) and hence rules out other possible locations.  The simulated STM
image, shown in Fig.~2(b), from a 5$\times$4 arrangement of Si adatoms
decorating these sites is in excellent agreement with the atomically
resolved images of
Ref.~[\onlinecite{bennewitz_nanotechnology_2002a}]. In particular, the
bright spots are correctly positioned in the middle of the underlying
row structures. The linescan in Fig.~2(c) shows their apparent height
to be 1.5 \AA,  in good agreement with the results of
Ref.~[\onlinecite{bennewitz_nanotechnology_2002a}].  Finally, Au
adatoms can be easily ruled out as plausible candidates: although
their binding energies are substantial (from 3.1 to 3.9 eV), they
relax well into the surface layer and thus produce no detectable STM
spot, as demonstrated in the inset to Fig.~2(b).

Photoemission data for Si(111)5$\times$2-Au provide a very stringent
test for any structural model. For wavevectors along the chain
direction, the ARPES data reveal a strong surface band beginning at
the 5$\times$2 zone boundary (the A$_2$ point), and dispersing
downward to its minimum at the 5$\times$1 zone boundary (the A$_1$
point) before turning back up \cite{altmann_phys_rev_b_2001a}. The
calculated electronic structure, shown in Fig.~3, reveals just such a
band between A$_2$ and A$_1$, whose width and effective mass (0.6 eV
and 0.4$m_e$) are in reasonably good agreement with the data (0.9 eV
and 0.5$m_e$). Additional bands with less pronounced surface character
are also found in the data, and can be tentatively identified with
calculated bands marked in Fig.~3.

\begin{figure}[tbp]
\resizebox{8.5cm}{!}{\includegraphics{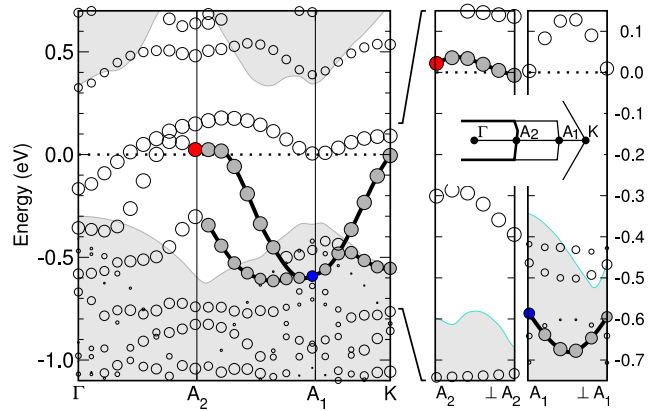}}
\caption{(color) Band structure of 5$\times$2 double honeycomb chain
model with no adatoms.  The size of each circle reflects
the surface character of the state (radii are proportional to the
total charge in spheres around surface-layer atoms). The shaded states
form bands that are detected in photoemission data (see
discussion). Left panel and right panels show dispersion along and
perpendicular to chain direction, respectively. The colored circles in
the left and right panels mark the same states; note the change of
energy scale.  }
\end{figure}

One remarkable feature of the ARPES results for the strong surface
band is the continuous transition, within a single band, from a
one-dimensional state (at the band maximum) to a two-dimensional state
(at its minimum) \cite{losio_phys_rev_lett_2000a}. This feature is
well reproduced in the calculated band structure, expanded views
of which are shown in the two small panels of Fig.~3 for wavevectors
perpendicular to the chain direction.  At the maximum of the strong
band, near A$_2$, the perpendicular dispersion is small (0.04 eV),
while at its minimum, near A$_1$, the perpendicular dispersion is much
larger (0.10 eV).  The ARPES measurements give very similar results,
0.03$\pm$0.03 eV and 0.14$\pm$0.03 eV, respectively
\cite{losio_phys_rev_lett_2000a}.

It is clear from Fig.~3 that the calculated electronic structure of
the bare Si(111)5$\times$2-Au reconstruction is metallic. As suggested
in Ref.~[\onlinecite{himpsel_journal_of_electron_spectroscopy_and_related_phenomena_2002a}],
it is interesting to ask whether the addition of Si adatoms could
render the system either insulating, or at least ``less metallic'';
whether it is energetically favorable to do so; and---if it
is---whether the predicted optimal coverage of adatoms corresponds to
the equilibrium coverage deduced from STM data, one per 5$\times$8
supercell
\cite{bennewitz_nanotechnology_2002a}. The answer to all three
questions is affirmative, as shown below.

Near the Fermi level the bands are very close to one-dimensional, as
discussed above. Hence, the number of additional electrons required to
render the band structure (provisionally assumed to be rigid)
insulating can be trivially determined from Fig.~3 to be two per
5$\times$2 cell.  To determine what coverage of Si adatoms is required
to provide two electrons per 5$\times$2 cell, explicit band structure
calculations must be performed.  For a coverage of one adatom per
5$\times$2 cell, the resulting band structure is insulating. This implies that the
addition of one adatom to the bare 5$\times$2 surface creates only one
additional state in the occupied manifold: two of the adatom's valence
electrons fill this state, and the other two dope the parent band
structure, rendering the full system insulating. These conclusions are
independent of any assumptions about the rigidity of the parent band
structure. Nevertheless, it is important to note that the bands do
shift quite rigidly, with only small changes of order 0.1 eV or less.
For lower adatom coverages, smaller but similarly rigid shifts are
expected.  This transforms the determination of the optimal adatom
coverage into the equivalent task of computing the optimal electron
doping.

In the absence of adatoms, the DFT surface energies of the 5$\times$1
and 5$\times$2 DHC models are very close, with the 5$\times$1 variant
preferred by less than 1 meV/\AA$^2$. Changes in the surface energy
due to electron doping were modeled by calculating the DFT total
energies for cells with additional electronic charge (with a
compensating background charge to preserve overall neutrality,
plus the standard correction to treat the resulting spurious
interactions \cite{makov_phys_rev_b_1995a}).  The results are shown in
Fig.~4. For the 5$\times$1 variant, any additional electronic charge
increases the surface energy above its undoped value.  The 5$\times$2
variant behaves quite differently: its surface energy is minimized for
a doping level very close to 0.5 electron per 5$\times$2 cell.  Since
each adatom was earlier shown to supply two doping electrons per
5$\times$2 cell, this optimal doping level can be most easily achieved
with one adatom per four 5$\times$2 cells.  Moreover, at this optimal
adatom coverage the surface energy of the 5$\times$2 variant is lower
than that of the 5$\times$1, suggesting that this is indeed the
observed ground-state equilibrium phase \cite{oneadatom}.  
These findings are in
excellent agreement with the experimental observation that when excess
Si is evaporated at low temperature onto a Si(111)5$\times$2-Au
surface with one adatom per four 5$\times$2 cells, subsequent
annealing at higher temperature will cause the extra Si to diffuse
away and return the system to its equilibrium state.

\begin{figure}[tbp]
\resizebox{8.0cm}{!}{\includegraphics{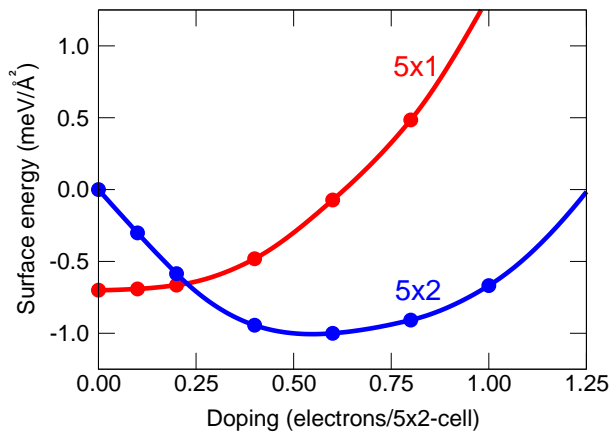}}
\caption{Variation of surface energy with electron doping, for
5$\times$1 and 5$\times$2 models. The most favorable structure has 5$\times$2
periodicity with 0.5 extra electron per 5$\times$2 cell.}
\end{figure}

In summary, a new structural model has been proposed for the
Si(111)5$\times$2-Au surface, consisting of a ``double honeycomb
chain'' reconstruction decorated by Si adatoms. Simulated STM images
from this model reproduce a number of experimentally observed
features, including the in-plane location and apparent height of the
bright protrusions due to the adatoms, and the ``Y''-shaped features
seen away from these protrusions. The calculated band structure
reproduces the main features of recent photoemission data, including
the unusual change in dimensionality observed within the main surface
band between its energy extrema. Finally, the Si adatoms
act as electron donors that dope the parent 5$\times$2
band structure, reducing its surface energy and stabilizing it
relative to other models.  The optimal doping level is equivalent to
one Si adatom per four 5$\times$2 cells, in agreement with
experimental observation.

\begin{acknowledgments}
Many helpful conversations with F.J. Himpsel are gratefully
acknowledged. Computations were performed at the DoD Major Shared
Resource Center at ASC. This work was supported by the Office of Naval
Research.
\end{acknowledgments}


\begin{thebibliography}{10}
\bibitem{oneadatom}
This analysis neglects other contributions to the energy, but these are
  expected to be small since the DFT surface energy changes by only 1
  meV/\AA$^2$ upon adding one adatom per 5$\times$2 cell.
\end{thebibliography}

\end{document}